\documentclass[manuscript,aps,prl]{revtex4}
\usepackage{epsfig}
\usepackage{graphicx}
\input{epsf}

\begin{document}

\title{Demonstration of Dispersion-Cancelled Quantum-Optical Coherence Tomography }

\author{Magued B. Nasr, Bahaa E. A. Saleh, Alexander V. Sergienko, and Malvin C. Teich\footnote{Email address: teich@bu.edu}}

\address{Quantum Imaging Laboratory\footnote{URL: http://www.bu.edu/qil}, Departments of Electrical $\&$ Computer Engineering and Physics, Boston University, Boston, MA 02215}

\begin{abstract}
We present an experimental demonstration of quantum optical
coherence tomography (QOCT). The technique makes use of an
entangled twin-photon light source to carry out axial optical
sectioning. QOCT is compared to conventional optical coherence
tomography (OCT). The immunity of QOCT to dispersion, as well as a
factor of two enhancement in resolution, are experimentally
demonstrated.

\end{abstract}

\maketitle

Optical coherence tomography (OCT) has become a versatile and
useful biological imaging technique
\cite{Fuj1,Fuj2,Schmidt-Review}, particularly in ophthalmology
\cite{Ophthalmology}, cardiology \cite{Cardiology}, and
dermatology \cite{Dermatology}. It is an interferometric scheme
that makes use of a light source of short coherence time (broad
spectrum) \cite{SalehTeich} to carry out axial sectioning of a
biological specimen. Axial resolution is enhanced by increasing
the spectral bandwidth of the source (sub-micrometer resolution
has recently been achieved by using a light source with a
bandwidth of 325 nm \cite{SubmicronRes}). However, as the
bandwidth is increased the effects of group-velocity dispersion
becomes increasingly deleterious \cite{DispersionEffects}. Various
techniques have been used in attempts to counteract the effects of
dispersion, but these require \textit{a priori} knowledge of the
dispersion intrinsic to the specimen \cite{DispersionCorrection1}.

A quantum version of OCT that makes use of an entangled
twin-photon light source has recently been proposed
\cite{Abouraddy}. A particular merit of quantum-optical coherence
tomography (QOCT) is that it is inherently immune to dispersion by
virtue of the frequency entanglement associated with the
twin-photon pairs \cite{Franson,Kwiat,Larchuk}. Moreover, for
sources of the same bandwidth, the entangled nature of the twin
photons provides a factor of two enhancement in resolution
relative to OCT.

In this letter we report the first experimental demonstration of
QOCT, and show that the technique is indeed insensitive to
group-velocity dispersion. A parallel experiment using
conventional OCT with a source of the same bandwidth is conducted
to provide a direct comparison of the two techniques. Using the
reflections from the two surfaces of a fused-silica sample buried
under a 10-mm-thick ZnSe window (a highly dispersive material) as
a sample, we obtain an improvement in resolution by a factor of
approximately 5. This improvement arises from the concatenation of
two effects: dispersion cancellation and the factor-of-two
advantage.

\begin{figure}[htbp]
  \begin{center}
    \epsfig{file=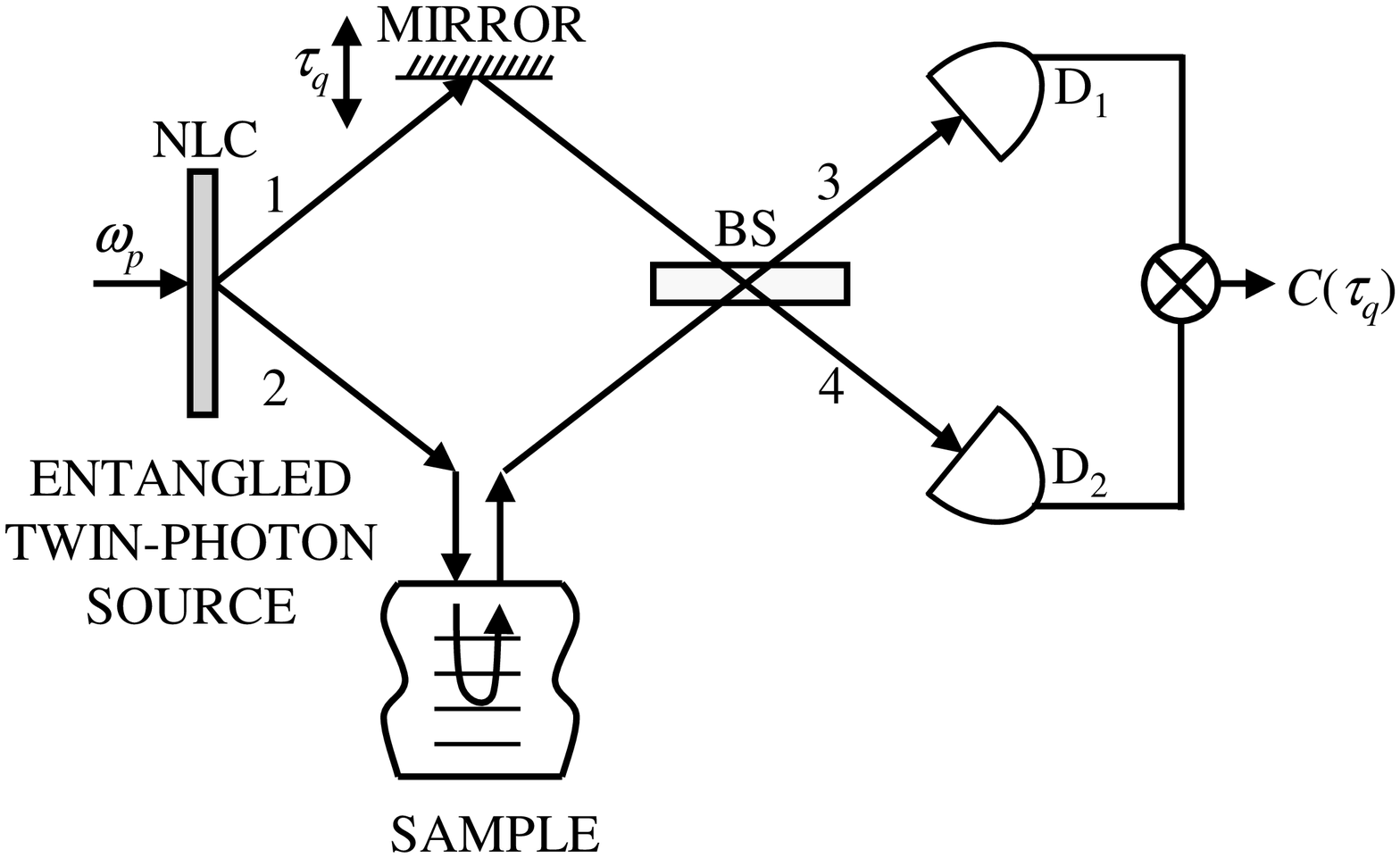,clip,width=6in}
    \caption{Schematic of quantum-optical coherence tomography (QOCT).
A monochromatic laser of angular frequency $\omega_{p}$ pumps a
nonlinear crystal (NLC), generating pairs of entangled photons. BS
stands for beam splitter and $\tau_{q}$ is an adjustable temporal
delay. D$_{1}$ and D$_{2}$ are single-photon-counting detectors
that feed a coincidence circuit indicated by the symbol
$\bigotimes$. The outcome of an experiment is the coincidence rate
$C(\tau_{q})$.}
    \label{SchematicQOCT}
  \end{center}
\end{figure}

We begin with a brief discussion of the principle underlying QOCT
(for a comparative review of the theories of QOCT and OCT the
reader is referred to Ref. \cite{Abouraddy}). A schematic of the
QOCT arrangement is illustrated in Fig. ~\ref{SchematicQOCT}. The
entangled twin photons may be conveniently generated via
spontaneous parametric down-conversion (SPDC) \cite{MandelWolf}.
In this process a monochromatic laser beam of angular frequency
$\omega_{p}$, serving as the pump, is sent to a second-order
nonlinear optical crystal (NLC). A fraction of the pump photons
disintegrate into pairs of downconverted photons. Both
downconveted photons have the same polarization and central
angular frequency $\omega_{0}\,=\,\omega_{p}/2$, corresponding to
type-I degenerate SPDC. We direct our attention to a noncollinear
configuration, in which the photons of the pairs are emitted in
selected different directions (modes), denoted 1 and 2. Although
each of the emitted photons has a broad spectrum in its own right,
the sum of the angular frequencies must always equal $\omega_{p}$
by virtue of energy conservation.

The twin-photon source is characterized by the frequency-entangled
state
\begin{equation}\label{FreqEntangled-state}
|\psi\rangle=\int d\Omega\,
\zeta(\Omega)\,|\omega_{0}+\Omega\rangle_{1} \,
|\omega_{0}-\Omega\rangle_{2},
\end{equation}
where $\Omega$ is the angular frequency deviation about the
central angular frequency $\omega_{0}$ of the twin-photon wave
packet, $\zeta(\Omega)$ is the spectral probability amplitude, and
the spectral distribution $S(\Omega)=|\zeta(\Omega)|^{2}$ is
normalized such that $\int d\Omega \, S(\Omega)=1$. For simplicity
we assume $S(\Omega)$ to be a symmetric function. Each photon of
the pair resides in a single spatial mode, indicated by the
subscripts 1 and 2 in Eq. (\ref{FreqEntangled-state}).

The schematic illustrated in Fig. \ref{SchematicQOCT} has, at its
heart, the two-photon interferometer considered by Hong, Ou, and
Mandel (HOM) \cite{HOM}. The conventional HOM configuration is
modified by placing the sample to be probed in one arm and an
adjustable temporal delay ($\tau_{q}$) in the other arm. The
entangled photons are directed to the two input ports of a
symmetric beam splitter (BS). Beams 3 and 4 at its output ports
are directed to two single-photon-counting detectors, D$_{1}$ and
D$_{2}$, respectively. The coincidence rate of photons arriving at
the two detectors, $C(\tau_{q})$, is recorded within a time window
determined by a coincidence circuit (indicated by $\otimes$).

An experiment is conducted by sweeping the temporal delay
$\tau_{q}$ and recording the interferogram $C(\tau_{q})$. If a
mirror were to replace the sample, this would trace out a dip in
the coincidence rate whose minimum would occur when arms 1 and 2
of the interferometer had equal path lengths. This dip would
result from interference of the two photon-pair probability
amplitudes, \textit{viz}. reflection or transmission of both
photons at the beam splitter.

For simplicity, we neglect losses in this exposition. A weakly
reflecting sample is then described by a transfer function
$H(\omega)$, characterizing the overall reflection from all
structures that comprise the sample, at angular frequency
$\omega$:
\begin{equation}\label{H(omeg)}
H(\omega)=\int_{0}^{\infty}dz \, r(z,\omega) \,
e^{i2\phi(z,\omega)}.
\end{equation}
The quantity $r(z,\omega)$ is the complex reflection coefficient
from depth $z$ and $2\,\phi(z,\omega)$ is the round-trip phase
accumulated by the wave while travelling through the sample to
depth $z$. As shown previously \cite{Abouraddy}, the coincidence
rate $C(\tau_{q})$ is then given by
\begin{equation}\label{Coincidence-Rate}
C(\tau_{q})\propto\Lambda_{0}-\textrm{Re} \,
\{\Lambda(2\tau_{q})\},
\end{equation}
where
\begin{equation}\label{BackGround}
\Lambda_{0}=\int d\Omega \, |H(\omega_{0}+\Omega)|^{2} \,
S(\Omega)
\end{equation}
and
\begin{equation}\label{Modulation}
\Lambda(\tau_{q})=\int d\Omega \, H(\omega_{0}+\Omega) \,
H^{\ast}(\omega_{0}-\Omega) \, S(\Omega) \, e^{-i\Omega\tau_{q}}
\end{equation}
represent the constant and varying contributions, respectively.
The interferogram $C(\tau_{q})$ yields useful information about
the transfer function $H(\omega)$ and hence about the reflectance
$r(z,\omega)$ \cite{Abouraddy}.

\begin{figure}[htbp]
  \begin{center}
    \epsfig{file=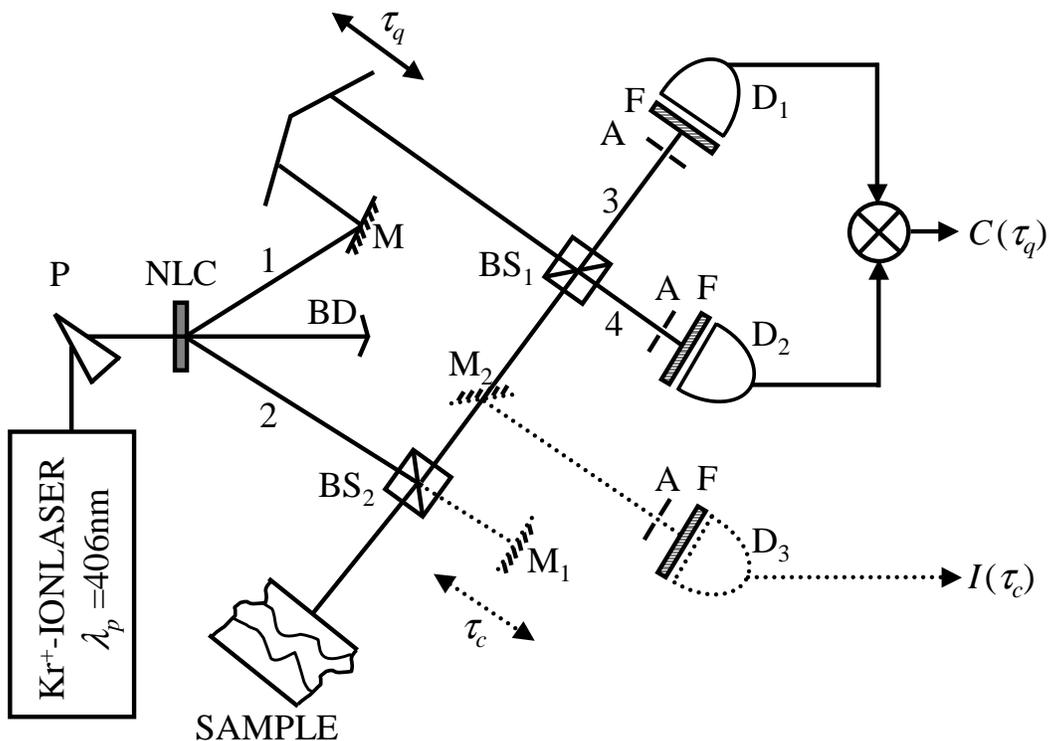,clip,width=6in}
\caption{Experimental arrangement for quantum$/$classical optical
coherence tomography QOCT/OCT. A monochromatic Kr$^{+}$-ion laser
operated at $\lambda_{p}$ = 406 nm pumps an 8-mm-thick type-I
LiIO$_{3}$ nonlinear crystal (NLC) after passage through a prism,
P, and an aperture (not shown), which remove the spontaneous glow
of the laser tube. BD stands for beam dump (to block the pump), BS
for beam splitter, M for mirror, A for 2.2-mm aperture, F for
long-pass filter with cutoff at 725 nm, and D for
single-photon-counting detector (EG$\&$G, SPCM-AQR-15). The
quantities $\tau_{q}$ and $\tau_{c}$ represent temporal delays.
For QOCT scans, the dotted components M$_{1}$, M$_{2}$ and D$_{3}$
are removed, $\tau_{q}$ is swept, and the coincidence rate
$C(\tau_{q})$ is measured within a 3.5-nsec time window. For OCT
scans, beam 1 is discarded (beam 2 serves as the
short-coherence-time light source), $\tau_{c}$ is swept, and the
singles rate $I(\tau_{c})$ is recorded.}
    \label{ExperimentalQOCT}
  \end{center}
\end{figure}

The details of the QOCT experimental arrangement are shown in Fig.
~\ref{ExperimentalQOCT}. For QOCT scans, the dotted components
(mirrors M$_{1}$ and M$_{2}$, as well as detector D$_{3}$) are
removed. The entangled photons, centered about $\lambda_{0}\,=
\,812\,\,\textrm{nm}$ and emitted in a non-collinear
configuration, travel in beams 1 and 2. The photon in beam 1
travels through a temporal delay $\tau_{q}$ before it enters the
input port of the first beam splitter, BS$_{1}$. The second photon
in beam 2 goes through a second beam splitter, BS$_{2}$, which
ensures normal incidence onto the sample. The photon returned from
the sample is directed to the other input port of BS$_{1}$. Beams
3 and 4, at the output of BS$_{1}$, are directed to D$_{1}$ and
D$_{2}$, respectively.

For OCT scans, the photons in beam 1 are discarded and mirrors
M$_{1}$ and M$_{2}$ remain in place (see Fig.
~\ref{ExperimentalQOCT}). The photons in beam 2 serve as a
short-coherence-time light source. The reflections from the sample
and mirror M$_{1}$, after recombination at beam splitter BS$_{2}$
are directed to detector D$_{3}$ via mirror M$_{2}$. The result is
a simple Michelson interferometer, the standard configuration for
OCT. To conduct an experiment, the temporal delay $\tau_{c}$ is
swept and the singles rate is recorded, forming the OCT
interferogram $I(\tau_{c})$. This arrangement permits a fair
comparison between QOCT and OCT since both make use of a light
source with identical spectrum.

The initial experiment makes use of a thin fused-silica window as
the sample. The transfer function $H(\omega)$ is then given by
\begin{equation}\label{2SurfTransFunct}
H(\omega)=r_{1}+r_{2}\,e^{i2\omega nL/c},
\end{equation}
where the reflectances from the front and back surfaces are
$|r_{1}|^{2}\,=\,|r_{2}|^{2}\,=\,0.04$ at normal incidence; $L$ =
90 $\mu$m is the sample thickness (which is greater than the
37-$\mu$m coherence length of the source), $c$ is the speed of
light in vacuum, and $n\,\approx\,1.5$ is the refractive index of
the fused silica (which is taken to be independent of $\omega$ by
virtue of the low dispersiveness of the material). Under these
conditions, Eqs. (\ref{BackGround}), (\ref{Modulation}), and
 (\ref{2SurfTransFunct}) yield
\begin{equation}\label{BackGround2Surf}
\Lambda_{0}=|r_{1}|^{2}+|r_{2}|^{2}
\end{equation}
and
\begin{equation}\label{Modulation2Surf}
\Lambda(\tau_{q})=|r_{1}|^{2}\,s(\tau_{q})+|r_{2}|^{2}\,s(\tau_{q}-2\tau_{d})+\,2\,\textrm{Re}\,\{r_{1}r_{2}^{\ast}\,s(\tau_{q}-\tau_{d})\,e^{i
\omega_{p}nL/c}\},
\end{equation}
where $s(\tau_{q})$ is the Fourier transform of the source
spectrum $S(\Omega)$. Substituting Eqs. (\ref{BackGround2Surf})
and (\ref{Modulation2Surf}) into Eq. (\ref{Coincidence-Rate})
yields an interferogram that contains the three varying terms in
Eq. (\ref{Modulation2Surf}).

The first two terms in Eq. (\ref{Modulation2Surf}) are dips
arising from reflections from each of the two surfaces. They are
separated by $\tau_{d}=2nL/c$ and are expected to exhibit 50$\%$
visibility since $|r_{1}|^{2}\,=\,|r_{2}|^{2}$. The third term,
which appears midway between the two dips, arises from
interference between the probability amplitudes associated with
these reflections. This term changes from a hump to a dip
depending on the values of $\omega_{p}$, $n$, $L$, and the
arguments of $r_{1}$ and $r_{2}$.

The experimental QOCT interferogram for this sample, normalized to
$\Lambda_{0}$, is plotted in Fig. 3a. The two dips, separated by
the optical path length of the sample $nL$ = 135 $\mu$m, exhibit
45$\%$ visibility, in close agreement with the theoretically
expected value of 50$\%$. The abscissa is represented in units of
the scaled temporal delay $c\tau/2$, representing the physical
displacement of the delay line, so that $\tau$ stands for
$\tau_{q}$ and $\tau_{c}$ alike.

The OCT interferogram for the same sample is expected to consist
of two interference-fringe envelopes, each with visibility
calculated to be 30$\%$, separated by $\tau_{d}$. The experimental
OCT interferogram for this sample, normalized to the constant
background, is shown in Fig. 3b. The centers of the envelopes,
separated by $nL$ = 135 $\mu$m, exhibit 28$\%$ visibility.

It is apparent that the 18.5-$\mu$m FWHM of the dips observed in
QOCT provides a factor of 2 improvement in resolution over the
37-$\mu$m FWHM of the envelopes observed in OCT. This improvement,
which is in accord with theory \cite{Abouraddy}, ultimately
results from the entanglement inherent in the nonclassical light
source used in QOCT.

To demonstrate the dispersion-cancellation capability of QOCT, we
bury the sample under a highly dispersive medium and carry out a
QOCT/OCT experiment, as described above. The transfer function of
this composite sample is then
$H_{\mathrm{disp}}(\omega)\,=\,H(\omega)\,\textrm{e}^{i2\beta(\omega)d}$,
where the transfer function of the fused-silica window $H(\omega)$
is given by Eq. (\ref{2SurfTransFunct}), $\beta(\omega)$ is the
wave number in the dispersive medium, and $d$ is the thickness of
the dispersive medium. We expand $\beta(\omega_{0}+\Omega)$ to
second order in $\Omega$:
$\beta(\omega_{0}+\Omega)\approx\beta_{0}+\beta'\Omega+\beta''\Omega^{2}$,
where $\beta'$ is the inverse of the group velocity $v_{0}$ at
$\omega_{0}$, and $\beta''$ represents group-velocity dispersion
(GVD) \cite{SalehTeich}.

Substituting $H_{\mathrm{disp}}(\omega)$ into Eq.
(\ref{Modulation}), the QOCT varying term for the buried sample
turns out to be
\begin{equation}\label{ModulationBuriedSample}
\Lambda_{\mathrm{disp}}(\tau_{q})=\Lambda(\tau_{q}-2\beta'd),
\end{equation}
which is simply a displaced version of the result obtained for the
sample in air, as provided in Eq. (\ref{Modulation2Surf}). Neither
$\beta_{0}$ nor the GVD parameter $\beta''$ appear in Eq.
(\ref{ModulationBuriedSample}); nor, in fact, do any higher
even-order terms. The cancellation of GVD is an important
signature of QOCT. In OCT $\beta''$ does not cancel and the result
is a degradation of depth resolution and a reduction of the
signal-to-noise ratio \cite{DispersionCorrection1}.

In the experimental realization, the fused-silica window is buried
beneath two cascaded 5-mm-thick windows of highly dispersive ZnSe,
slightly canted with respect to the incident beam to divert
back-reflections, as shown at the top of Fig. 4. The GVD
coefficient for ZnSe is $\beta''=5\times10^{-25}\,$s$^{2}$m$^{-1}$
at $\lambda_{0}=\,812$ nm, which is 25 times greater than that for
fused silica.

Figure 4a shows the normalized QOCT interferogram for the buried
sample. As predicted in Eq. (\ref{ModulationBuriedSample}), the
widths of the dips remain 18.5 $\mu$m, just as they were in the
absence of the dispersive medium (see Fig. 3a).\textit{ Thus the
resolution of the QOCT scan is \textit{unaffected} by the presence
of the dispersive medium}. The hump between the two dips in the
QOCT interferogram in Fig. 4a is also unaffected by the presence
of the dispersive medium. This would not be the case, however, if
the dispersive material were between the reflecting surfaces
rather than outside of them \cite{Abouraddy}. On the other hand,
the interference-fringe envelopes in the normalized OCT
interferogram displayed in Fig. 4b are \textit{broadened} from 37
to 92 $\mu$m as a result of dispersion.

It is also of interest to compare the visibility of the
interferograms. Diverted reflection losses are not expected to
reduce the visibility of the QOCT features, whereas this benefit
does not accrue to OCT \cite{Abouraddy}. The robustness of QOCT in
this connection is evident in Fig. 4a, where the reduction of
visibility (in comparison with Fig. 3a) arises only from
misalignment. In Fig. 4b, on the other hand, the OCT interferogram
suffers a substantially greater loss of visibility (in comparison
with Fig. 3b), as a result of both dispersion and diverted losses
from the four uncoated ZnSe surfaces.

In conclusion, we have carried out proof-of-principle experiments
demonstrating the successful operation of a new axial optical
sectioning technique, quantum optical coherence tomography (QOCT).
We have experimentally demonstrated the two principal advantages
that stem from the frequency entanglement of the twin-photon
source: dispersion cancellation and resolution doubling.

This work was supported by the National Science Foundation; by the
Center for Subsurface Sensing and Imaging Systems (CenSSIS), an
NSF Engineering Research Center; and by the David $\&$ Lucile
Packard Foundation. We are grateful to G. Di Giuseppe, A.
Abouraddy, and M. Hendrych for assistance with the experiments and
helpful comments.

\newpage

\newpage

\newpage

\begin{figure}[htbp]
  \begin{center}
    \epsfig{file=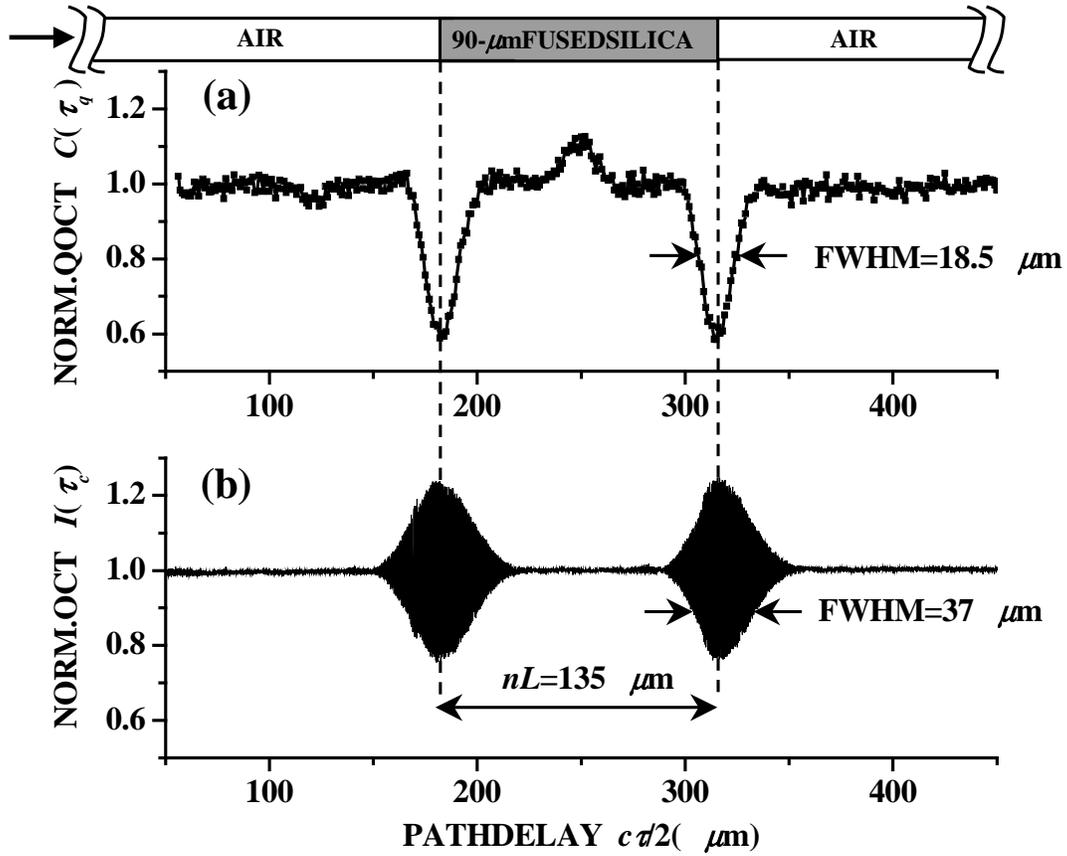,clip,width=6in}
\caption{QOCT and OCT normalized interferograms for a 90-$\mu$m
fused-silica window in air (as shown at top of figure). The
abscissa is the scaled temporal delay $c\tau/2$, which represents
displacement of the delay line ($\tau$ therefore represents both
$\tau_{q}$ and $\tau_{c}$). (a) Coincidence rate $C(\tau_{q})$
normalized to $\Lambda_{0}$ (the QOCT normalized interferogram).
(b) Singles rate $I(\tau_{c})$ normalized to constant background
(the normalized OCT interferogram).}
    \label{UndispData}
  \end{center}
\end{figure}

\newpage

\begin{figure}[htbp]
  \begin{center}
    \epsfig{file=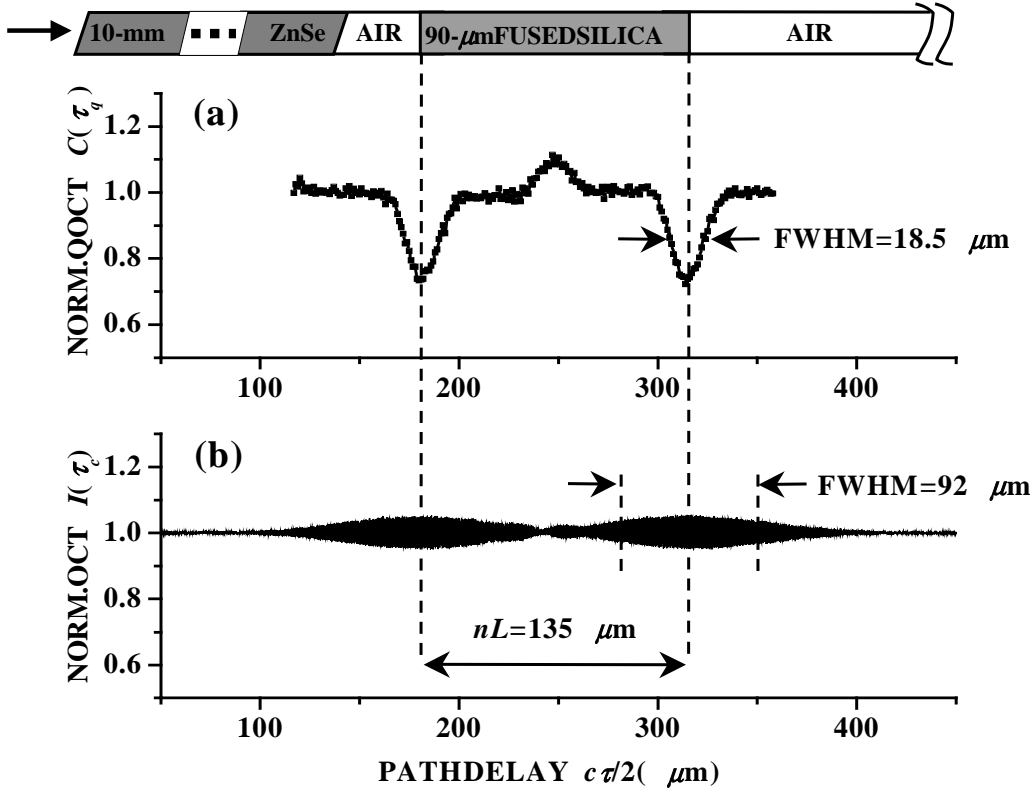,clip,width=6in}
\caption{QOCT and OCT normalized interferograms for a 90-$\mu$m
fused-silica window buried beneath two cascaded 5-mm-thick windows
of highly dispersive ZnSe. As shown at the top of the figure, the
ZnSe is slightly canted with respect to the incident beam (arrow)
to divert back-reflections. The abscissa is the scaled temporal
delay $c\tau/2$, which represents displacement of the delay line
($\tau$ therefore represents both $\tau_{q}$ and $\tau_{c}$). (a)
Coincidence rate $C(\tau_{q})$ normalized to $\Lambda_{0}$ (the
QOCT normalized interferogram). (b) Singles rate $I(\tau_{c})$
normalized to constant background (the normalized OCT
interferogram).}
    \label{DispData}
  \end{center}
\end{figure}

\end{document}